# Analysis of Two Time Scale Property of Singularly Perturbed System on Chaotic Attractor

**Mozhgan Mombeini, Ali Khaki Sedigh, Mohammad Ali Nekoui**

Science and Research Branch, Islamic Azad University, Hesarak, Punak, Tehran, Iran
E-mail: M.Mombeini80@gmail.com, Sedigh@kntu.ac.ir, Manekoui@eetd.kntu.ac.ir

**Abstract**
The idea that chaos could be a useful tool for analyze nonlinear systems considered in this paper and for the first time the two time scale property of singularly perturbed systems is analyzed on chaotic attractor. The general idea introduced here is that the chaotic systems have orderly strange attractors in phase space and this orderly of the chaotic systems in subscription with other classes of systems can be used in analyses. Here the singularly perturbed systems are subscripted with chaotic systems.
Two time scale property of system is addressed. Orderly of the chaotic attractor is used to analyze two time scale behavior in phase plane.

## 1. Introduction

Phase space analysis is common method in analysis of nonlinear systems [3]. Chaotic systems are class of nonlinear systems that are known by dependance of system dynamics on initial values. Since for first time in 1963 chaotic property introduced by Lornz , many researchers have shown interest in the analysis of them. On other hand nonlinear Singular perturbation models are known by dependence of the system properties on the perturbation parameter [3]. Multiple time scale characteristic is an important property of this class of systems. In this paper for the first time the two time scale property of singularly perturbed systems is analyzed on chaotic attractor. The general idea introduced here is that the chaotic systems have orderly strange attractors in phase space and this orderly of the chaotic systems in subscription with other classes of systems can be used in analyses. Here the singularly perturbed systems are subscripted with chaotic systems.

Two time scale property of system is addressed. Orderly of the chaotic attractor is used to analyze two time scale behavior in phase plane. Linearization method only gives the information around the point that system is linearized but phase space analysis gives all information about all points of the system. Mathematical models of ecological systems are examples of chaotic sinularly perturbed systems that analysis done on them here.

The paper is organized as follows. In section 2 the two time scale property of the singularly perturbed systems is introduced. In section 3 linearization method introduced to analyze the time scale.

Section 4 presents the results of employing the linearization method on the three ecological prey-predator systems.

In section 5 the two time scale behavior of singularly perturbed system on the chaotic attractor is analyzed. Section 6 contains the conclusion of this paper.

## 2. Two Time Scale Singularly Perturbed Systems

In this paper, chaotic singularly perturbed systems of the following form are considered,

$$\varepsilon\dot{x} = f(x,y)$$
$$\dot{y} = g(x,y) \quad \textbf{(1)}$$

Where, $x \in R$ , $y \in R^{n-1}$ and $\varepsilon$ is a small parameter. $f : R^n \to R$, $g : R^n \to R^{n-1}$ are both smooth functions and the system is chaotic.

The slow manifold of (1) is defined with

$$0 = f(x,y)$$
$$\dot{y} = g(x,y) \quad (2)$$

This $S$ manifold $S : \{f = 0\}$ is smooth and results in separation of time scales as $x$ the fast, and $y$ as the slow variable. It is easily seen that,

$$\frac{\dot{y}}{\dot{x}} = \frac{g(x,y)}{\varepsilon f(x,y)} \xrightarrow{\varepsilon \langle\langle 1} \frac{\dot{y}}{\dot{x}} \propto \frac{1}{\varepsilon} \quad (3)$$

By taking

$$\tau = \frac{t}{\varepsilon} \quad (4)$$

as the slow time and $t$ as the fast time, rescaling gives

$$\frac{dx}{d\tau} = x' = f(x,y)$$

$$\frac{dy}{d\tau} = y' = \varepsilon g(x,y)$$

The fast manifold yields:

$$x' = f(x,y)$$

$$y' = 0$$

**3. Analysis of Two Time Scale Behavior with Linearization around Slow Manifold**

In this section system (1) is linearized around its fixed point. Then slow manifold produced with (2). Then eigenvalues of jacobian matrix for full system and reduced system (slow manifold) used to analyze the speed of states. The equations

$$\dot{x} = 0$$
$$\dot{y} = 0 \qquad (5)$$

or equivalently the equations

$$0 = f(x,y)$$

$$0 = g(x,y) \quad (6)$$

give the fixed points $(x_{eq}, y_{eq})$ of the system (1). And according to (2) the slow manifold yields with

$$S = \{(y) : x = x_{eq}\} \quad (7)$$

Linearization of full system around the fixed point result in fallowing jacobian matrix

$$J = \begin{bmatrix} \frac{1}{\varepsilon} \times \frac{\partial f}{\partial x} & \cdots & \cdots & \frac{1}{\varepsilon} \times \frac{\partial f}{\partial y_{n-1}} \\ \frac{\partial g_1}{\partial x} & \frac{\partial g_1}{\partial y_1} & \cdots & \frac{\partial g_1}{\partial y_{n-1}} \\ \vdots & \cdots & \cdots & \vdots \\ \frac{\partial g_{n-1}}{\partial x} & \frac{\partial g_{n-1}}{\partial y_1} & \cdots & \frac{\partial g_{n-1}}{\partial y_{n-1}} \end{bmatrix} \quad (8)$$

and linearization of reduced system result in following jacobian matrix

$$J = \begin{bmatrix} \frac{\partial g_1}{\partial y_1} & \cdots & \frac{\partial g_1}{\partial y_{n-1}} \\ \vdots & \cdots & \vdots \\ \frac{\partial g_{n-1}}{\partial y_1} & \cdots & \frac{\partial g_{n-1}}{\partial y_{n-1}} \end{bmatrix} \quad (9)$$

It is obvious that eigenvalues with nonzero real parts of this matrixes (8),(9) show the speeds of states around the fixed pointes.

**4. The Linearization Method on Three Ecological Models**

Here the linearization method is implemented on three models of food chains of prey-predator type. The Rosenzweig–MacArthur, the Hastings–Powell, and the Volterra–Gause model are investigated here. All are singularly perturbed and the chaotic property of them in some range of parameters is proved in [1-2].The models include three states: a prey ($x$), a predator ($y$) and a top-predator ($z$).

**4.1. The Rosenzweig–Mac Arthur Model**

$$\varepsilon\frac{dx}{dt} = x(1 - x - \frac{y}{\beta_1 + x})$$
$$\frac{dy}{dt} = y(\frac{x}{\beta_2 + x} - \delta_1 - \frac{z}{\beta_2 + y}) \quad (10)$$
$$\frac{dz}{dt} = \xi z(\frac{y}{\beta_2 + y} - \delta_2)$$

Where

$$\beta_1 = 0.3, \beta_2 = 0.1, \delta_1 = 0.1, \delta_2 = 0.62, \xi = 0.3$$

Fixed point $(0.8593, 0.1632, 0.1678)$ is on the slow manifold. Eigenvalues of Jacobian matrix around this point for full system and reduced system for $\varepsilon = 0.1$ are

$$\lambda(\varepsilon = 0.1) = \begin{bmatrix} -7.516 \\ 0.182 + 0.112i \\ 0.182 - 0.112i \end{bmatrix}$$

$$\lambda_{reduced}(\varepsilon = 0.1) = \begin{bmatrix} 0.199 + 0.0759i \\ 0.199 - 0.0759i \end{bmatrix}$$

For $\varepsilon = 0.01$ eigenvalues change to

$$\lambda(\varepsilon = 0.01) = \begin{bmatrix} -75.467 \\ 0.181 + 0.112i \\ 0.181 - 0.112i \end{bmatrix}$$

$$\lambda_{reduced}(\varepsilon = 0.01) = \begin{bmatrix} 0.199 + 0.076i \\ 0.199 - 0.076i \end{bmatrix}$$

**4.2. The Volterra–Gause Model**

$$\varepsilon\frac{dx}{dt} = x(1-x) - \sqrt{x}y$$
$$\frac{dy}{dt} = -\delta_1 y + \sqrt{x}y - \sqrt{y}z \qquad \textbf{(11)}$$
$$\frac{dz}{dt} = \xi z(\sqrt{y} - \delta_2)z$$

Where

$$\delta_1 = 0.577, \delta_2 = 0.376, \xi = 1.428$$

Fixed point $(0.8463235, 0.141376, 0.1289524)$ is on the slow manifold. Eigenvalues of jacobian matrix around this point for full system and reduced system for $\varepsilon = 0.1$ are

$$\lambda(\varepsilon = 0.1) = \begin{bmatrix} -7.604 \\ 0.040 + 0.302i \\ 0.040 - 0.302i \end{bmatrix}$$

$$\lambda_{reduced}(\varepsilon = 0.1) = \begin{bmatrix} 0.086 + 0.291i \\ 0.086 - 0.291i \end{bmatrix}$$

For $\varepsilon = 0.01$ eigenvalues change to

$$\lambda(\varepsilon = 0.01) = \begin{bmatrix} -76.857 \\ 0.040 + 0.301i \\ 0.040 - 0.301i \end{bmatrix}$$

$$\lambda_{reduced}(\varepsilon = 0.01) = \begin{bmatrix} 0.086 + 0.291i \\ 0.086 - 0.291i \end{bmatrix}$$

**4.3. The Hastings–Powell Model**

$$\varepsilon\frac{dx}{dt} = x(1-x) - \frac{a_1 xy}{1+\beta_1 x})$$
$$\frac{dy}{dt} = y(\frac{a_1 x}{1+\beta_1 x} - \delta_1) - \frac{a_2 yz}{1+\beta_2 y}) \quad (12)$$
$$\frac{dz}{dt} = \xi z(\frac{a_2 y}{1+\beta_2 y} - \delta_2)$$

Where

$$a_1 = 5, a_2 = 0.1, \delta_1 = 0.4, \delta_2 = 0.01, \beta_1 = 3, \beta_2 = 2, \xi = 0.2$$

Fixed point $(0.8192, 0.125, 9.808)$ is on the slow manifold. Eigenvalues of Jacobian matrix around this point for full system and reduced system for $\varepsilon = 0.1$ are

$$\lambda(\varepsilon = 0.1) = \begin{bmatrix} -6.818 \\ 0.034 + 0.011i \\ 0.034 - 0.101i \end{bmatrix}$$

$$\lambda_{reduced}(\varepsilon = 0.1) = \begin{bmatrix} 0.148 \\ 0.008 \end{bmatrix}$$

For $\varepsilon = 0.01$ eigenvalues change to

$$\lambda(\varepsilon = 0.01) = \begin{bmatrix} -68.978 \\ 0.034 + 0.011i \\ 0.034 - 0.101i \end{bmatrix}$$

$$\lambda_{reduced}(\varepsilon = 0.01) = \begin{bmatrix} 0.148 \\ 0.008 \end{bmatrix}$$

It is obvious that for three systems fast mode is in perturbed direction. Fast modes are stable and very big in comparison to other poles. With decrement of $\varepsilon$ stable fast mode becomes faster and slow modes approximately remain unchanged. Then two time scale behavior in such systems means that with decrement of $\varepsilon$ value fast states become faster and slow sates speed is approximately unchanged. The eigenvalues of the reduced system (slow manifold) also remain unchanged with $\varepsilon$ value changes.

**5. Phase Space Analysis on Chaotic Attractor**

Phase space analysis is common method in analysis of nonlinear systems. For nonlinear systems the phase portrait of a solution is a plot in phase space of the orbit evolution [4]. One of the most important properties of chaotic systems is that they have strange attractors; that has an apparent qualitative and bounded shape for each systems in range of parameters that system is chaotic and initial conditions that arisen from basin of attraction. We named here this property as orderly of the chaotic systems.

Strange attractor can be shown with plot of trajectories in phase portrait. Here the property of chaotic systems that "qualitative shape of system attractor is unchanged and bounded", or in other expression the orderly of the strange attractor of chaotic systems in phase portrait, is used to analyze the two time scale behavior of singularly perturbed chaotic systems in phase space.

According to (3) by $\varepsilon$ variation, speeds of systems states meet different scale times proportional to $1/\varepsilon$, theoretically. Figures (1) shows the strange attractor of three above ecological systems for two different $\varepsilon$ values in phase space. According to figure (1) by changing the $\varepsilon$ value the qualitative shape of attractor is approximately unchanged.
Figure (2) shows the two dimensional plot of attractors for fast states (y,z). According to figure (2) the qualitative shape and quantitative domain of variation of attractor for slow states is approximately independent on variation of $\varepsilon$ value and there is no sensible variation in slow states.

Figure (3) shows the two dimensional plot of the same attractors for the fast state (x) and one of the slow states(y). It shows that the speed of states increase in fast direction.

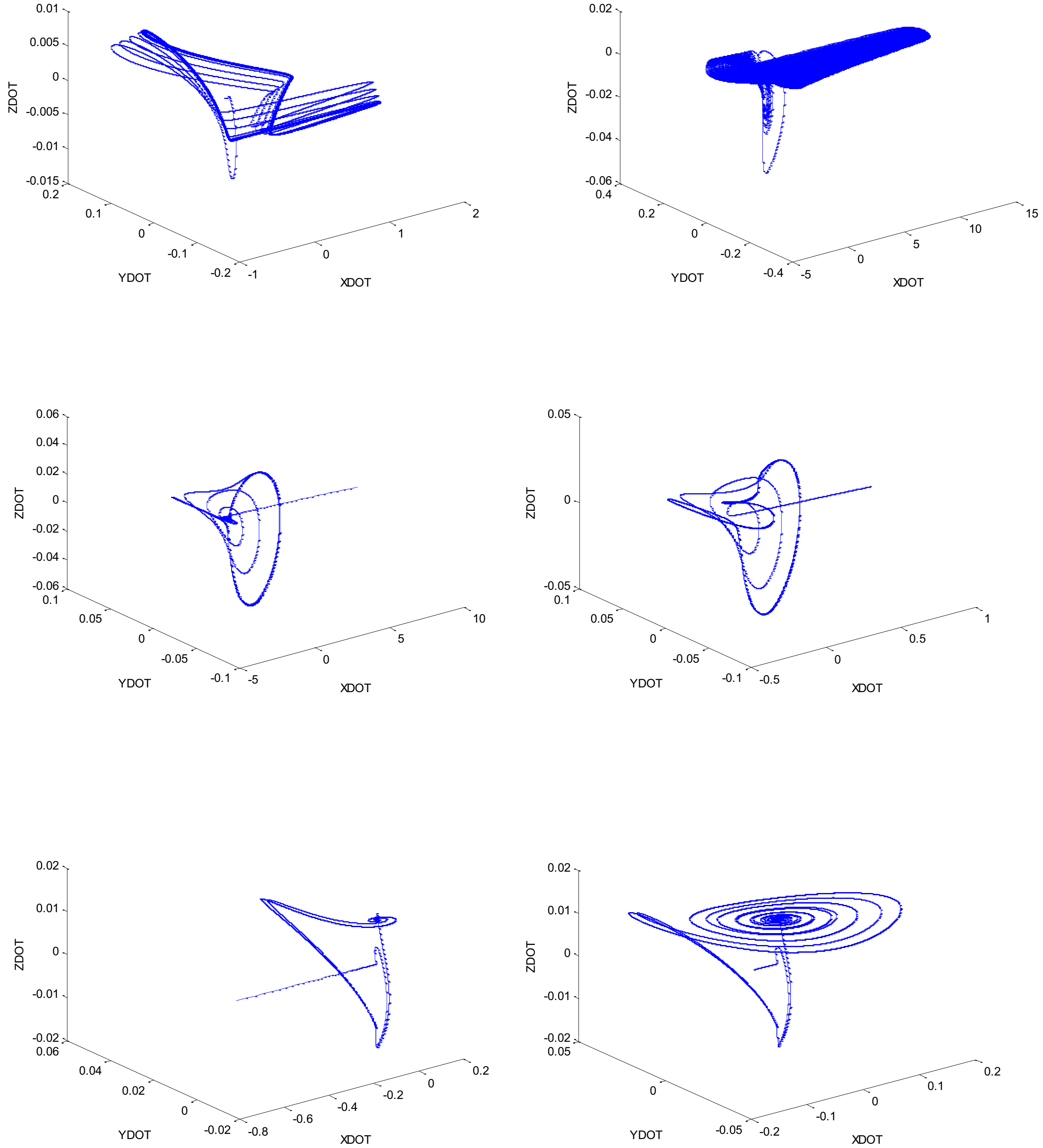

Figure (1) chaotic strange attractor of three food chain models (for $\varepsilon = 0.1$ in left and for $\varepsilon = 0.01$ in right).

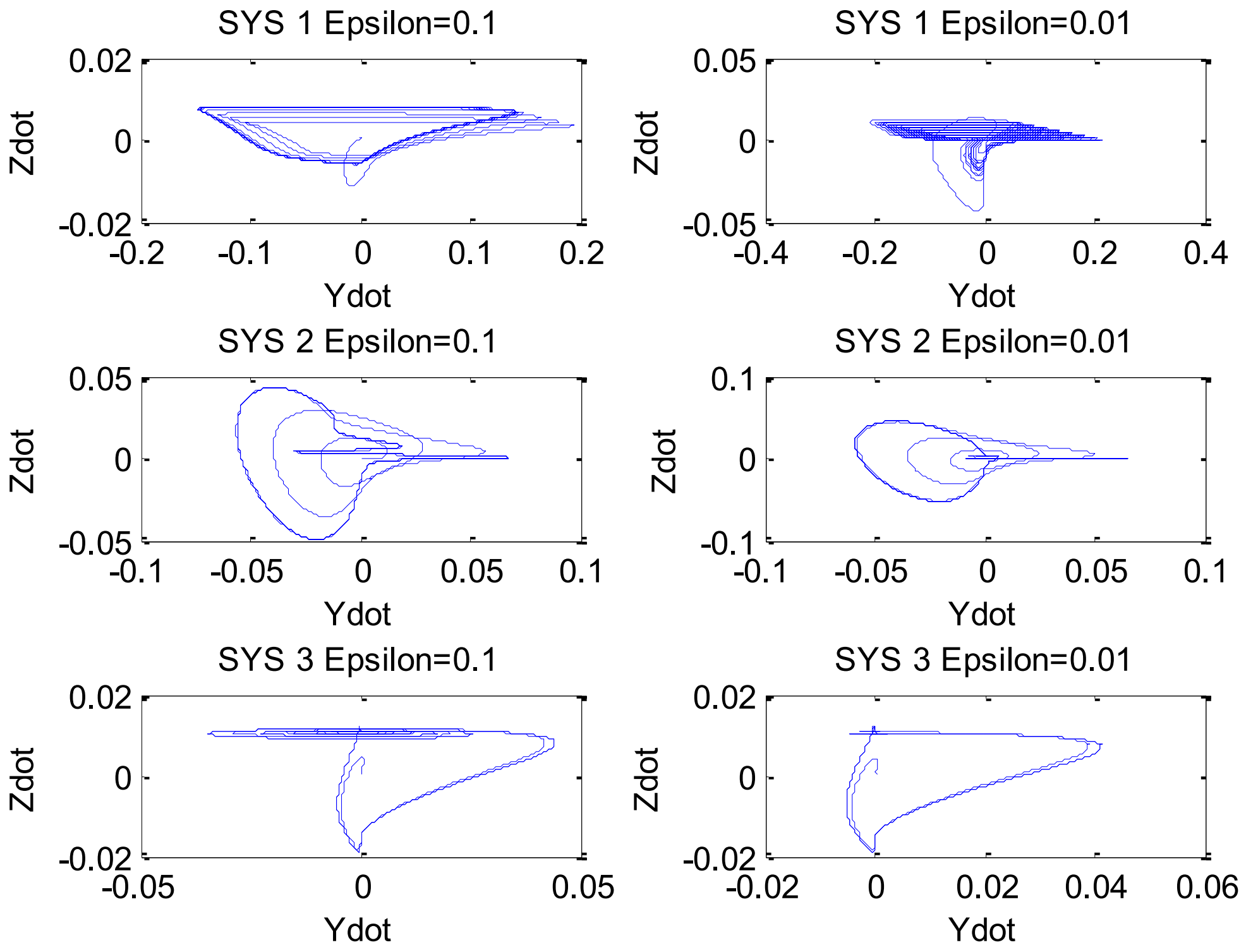

Figure (2) 2-Dimensional perspective of chaotic attractor of three food chains models for slow states (for $\varepsilon = 0.1$ in left and for $\varepsilon = 0.01$ in right).

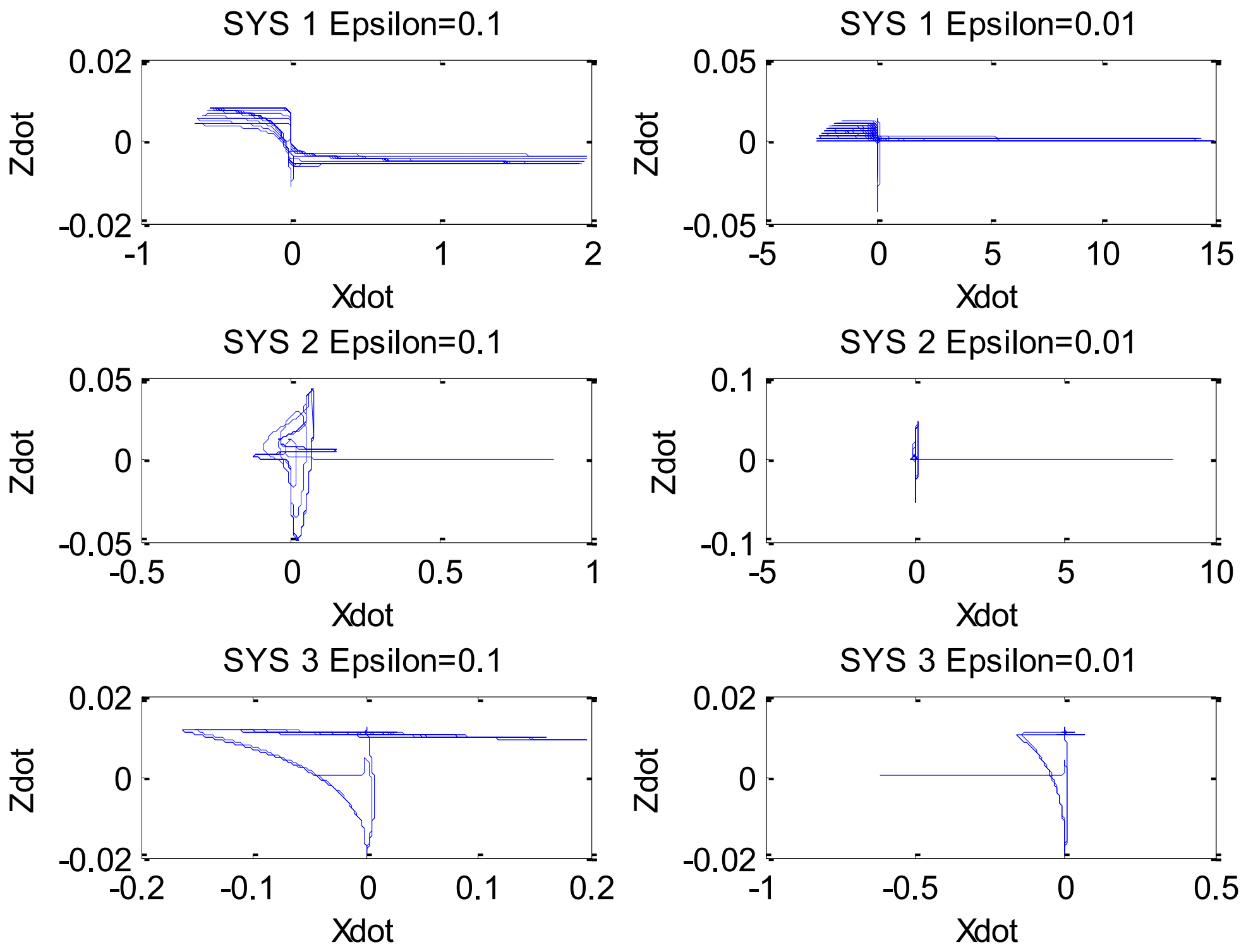

Figure (3) 2-Dimensional perspective of chaotic attractor of three food chains models for fast state $x$ (respect to one of the slow states $z$ (for $\varepsilon = 0.1$ in left and for $\varepsilon = 0.01$ in right).

According to these figures, the quantitative domain of variation of attractor in the direction of fast state increased by the decrement of $\varepsilon$ value, and for slow states is approximately is no sensible variation.
Then, analyze of the two scale time behavior of singularly perturbed systems on chaotic attractor shows that be $\varepsilon$ decrement the slow states speed is approximately unchanged but the speed of fast states increase. This result is for all trajectories of the system not only about the around the fixed point on slow manifold.

## 6. Conclusions

In linearization method the eigenvalues with nonzero real parts introduced to analyze the multi time scale property of system around the point that system linearized. Results of implementation of this method on three ecological models show that the eigenvalues of jacobian matrix in fast direction are very bigger than slow directions. To analyze the system behavior on all points the phase portrait method is used. Because the system is chaotic its strange attractor in phase portrait is bounded and has a regular qualitative shape. Using phase portrait method satisfied the results of linearization method but applicable for all points of the system on the strange attractor. Both method show that by decrement of $\varepsilon$ value the speed of fast state increases but the speed of slow states are approximately unchanged. The orderly of the chaotic system on strange attractor used to analyze the two scale time property of the singularly perturbed class of nonlinear systems. Using chaotic property in subscription with other classes of nonlinear systems may be extendable to analyze them.

## Refrences

[1] B. Deng. Food chain chaos due to junction fold point. Am. Inst. Phys 11:514–525, 2001.

[2] JM. Ginoux, B. Rossetto and JL. Jamet. Chaos in a three-dimensional Volterra-Gause model of predator-prey type. International Journal of Bifurcation and Chaos Vol. 15, No. 5: 1689-1708, 2005.

[3] H. Khalil. Nonlinear systems. Michihan state University [2nd ed.], 1996.

[4] K. Tomasz and S. R. Bishop. The illustrated dictionary of nonlinear dynamics and chaos. John wiley & sons,1999.